\documentclass[english,10pt,aps,prb,twocolumn,notitlepage,superscriptaddress] {revtex4-1}

\usepackage{graphicx}
\usepackage{amssymb}
\usepackage{amsmath}
\usepackage{setspace}
\usepackage{babel}
\usepackage{epstopdf}
\usepackage{color}

\definecolor{OliveGreen}{rgb}{0,0.6,0}

\begin{document}
\title{Single layer MoS$_2$ nanoribbon field effect transistor}

\author{D. Kotekar-Patil}
\email[]{patild@imre.a-star.edu.sg}
\affiliation{Institute of Materials Research and Engineering, Agency for Science Technology and Research, 2 Fusionopolis Way, 08-03 Innovis, 138634, Singapore}
\author{J. Deng}
\affiliation{Institute of Materials Research and Engineering, Agency for Science Technology and Research, 2 Fusionopolis Way, 08-03 Innovis, 138634, Singapore}
\author{S. L. Wong}
 \affiliation{Institute of Materials Research and Engineering, Agency for Science Technology and Research, 2 Fusionopolis Way, 08-03 Innovis, 138634, Singapore}
 \affiliation{Department of Physics, National University of Singapore, 2 Science Drive 3, 117551, Singapore}
 \author{Chit Siong Lau}
 \affiliation{Institute of Materials Research and Engineering, Agency for Science Technology and Research, 2 Fusionopolis Way, 08-03 Innovis, 138634, Singapore}
\author{Kuan Eng Johnson Goh}
\email[]{kejgoh@yahoo.com}
\affiliation{Institute of Materials Research and Engineering, Agency for Science Technology and Research, 2 Fusionopolis Way, 08-03 Innovis, 138634, Singapore}
\affiliation{Department of Physics, National University of Singapore, 2 Science Drive 3,  117551, Singapore}

\begin{abstract}

We study field effect transistor characteristics in etched single layer MoS$_2$ nanoribbon devices of width 50\,nm with ohmic contacts. We employ a SF$_6$ dry plasma process to etch MoS$_2$ nanoribbons using low etching (RF) power allowing very good control over etching rate. Transconductance measurements reveal a steep sub-threshold slope of 3.5V/dec using a global backgate. Moreover, we measure a high current density of 38\,$\mu$A/$\mu$m resulting in high on/off ratio of the order of 10$^5$. We observe mobility reaching as high as 50\,cm$^2$/V.s with increasing source-drain bias. 
\end{abstract}

\maketitle

 
2D layered materials such as graphene and transition metal dichalcogenides (TMDs) provide an alternative path towards future nanoscale devices for electronic applications due to their interesting electronic properties. 
Graphene has been reported to exhibit very high charge mobility \citep{Chen2007}, but the lack of bandgap in graphene makes it unattractive for transistor applications \citep{Geim2005} . 
Bulk TMDs like MoS$_2$, WS$_2$,WSe$_2$ etc. are semiconducting with indirect bandgap whereas monolayer has a direct and sufficiently large bandgap (1-2eV) \citep{Heinz2010, Tongay2014, Qihua2014, Eda2013, Zhang2014, Baugher2014} to allow on and off transistor operations. 
Among TMD materials, MoS$_2$ is one of the well explored and broadly studied materials.
Additionally, MoS$_2$ has been reported to exhibit high mobility \citep{Liu2018, Pisoni2018} , high on/off ratio \citep{Radisavljevic2011} along with low contact resistance \citep{Liu2013,Yang2014, Liu2016} and ultrashort channel \citep{Desai2016} makeing it an attractive platform for transistor based applications.
The ultrathin body of single layer MoS$_2$ is advantageous for scaling the channel length below 5\,nm \citep{Daniel2018}.

In order to implement large scale TMD-based devices using top-down engineering, there are two initial challenges to be overcome:

1. Controlled growth of wafer scale TMDs, with specified number of layer(s) and homogeneity. 

2. Controlled etching of TMDs in order to define the active device region, e.g. nanoribbons (NR) which defines a conduction channel.

While advancements in large scale TMD growth is evidenced by the increasing body of works reported \citep{Manzeli2017} , relatively less focussed on achieving controlled TMD etching down to the few nm scale with reproducible specifications.
Etching can introduce edge roughness and defects which can deteriorate the charge transport in the NR by scattering charge carriers.
This is exacerbated in single layer MoS$_2$ due to the lack of screening in its ultrathin body.

In this manuscript, we investigate field effect transistor (FET) characteristics of 50\,nm wide etched NRs.
NRs are etched out using a dry SF$_6$ plasma.
Robust ohmic contacts are formed by maximizing the monolayer MoS$_2$ and metal contact overlap.
FET characterization is done by utilizing the global back gate.
The associated transfer characteristics reveal high field effect mobility of 50\,cm$^2$/V.s, current on/off ratio of 10$^5$ and current density (J) 38\,$\mu$A/$\mu$m demonstrating the highest values reported for TMD NR-FET reported to date for etched single layer MoS$_2$.

In order to fabricate a NR-FET, a MoS$_2$ flake was mechanically exfoliated from bulk crystal (2D semiconductors) and transferred on heavily p-doped Si substrate with a 300 nm thick SiO$_2$ layer on top. 
The substrate was also used as a back gate to investigate FET characteristics. 
The substrate with MoS$_2$ flake was then spin-coated with PMMA A5 and e-beam lithography was performed to define source and drain contacts (fig.1a).
Ti/Au (10/80\,nm) was evaporated followed by a lift-off step to form ohmic contacts with the MoS$_2$ (fig.1b).
The sample was then annealed at 200 degrees celsius in H$_2$/Ar atmosphere to ensure good ohmic contact and removal of any remaining resist residue \citep{Ishigami2007, Radisavljevic2011} .
In order to etch out a NR, the flake was spin-coated again with PMMA A5 (1:1)(fig.1c).
This time a thinner PMMA was used to get better e-beam lithographic resolution.
E-beam lithography was performed to write a structure such that in the region where NR was desired, PMMA remained unexposed to the e-beam.
This PMMA served as a mask to protect MoS$_2$ underneath it during etching process while the surrounding MoS$_2$ was etched (fig.1d).
After the etching process, the PMMA was removed with acetone leaving a MoS$_2$-NR between the metallic source-drain contacts (fig.1e).

Multilayer flakes were used to determine the optimal MoS$_2$ etching rates.
An SF$_6$ dry plasma was used to etch MoS$_2$  (SF$_6$: 30\,sccm, Pressure: 30\,mT, RF source power: 15\, W).
In order to achieve slower etching rate, we used a reduced RF source power as compared to previous studies\citep{Xiao2016, Zhang2018} . 
This gave better control over the etching rate and possibly minimizes the edge roughness.
With these etching parameters, we estimate an etching rate of 0.2 layer/sec.
For monolayer MoS$_2$-NR devices we always over-etched to ensure complete removal of MoS$_2$ in the desired area.
Fig.1f and 1g shows a series of optimised  etched NRs in multi-layer MoS2 with widths ranging from 50 nm to 185 nm.

\begin{figure}
\centering\includegraphics[width=1\columnwidth]{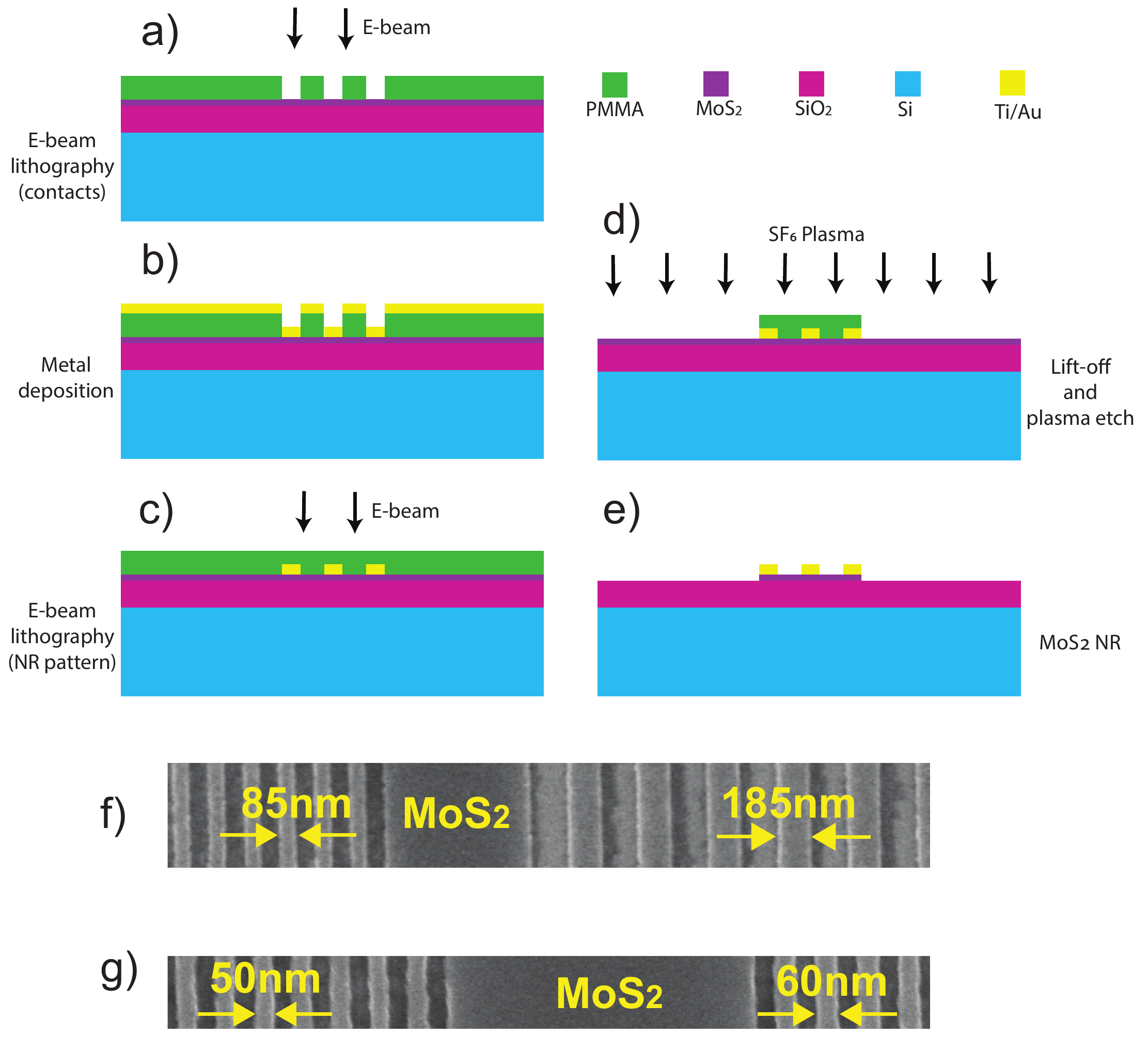}
\caption{Panels in figs. a-e shows a step-by-step process flow used to etch out a NR from MoS$_2$ flake. Two e-beam lithography steps are done to fabricate a NR-FET, first to define source and drain contacts and second to define NR structure in PMMA. SF$_6$ plasma was used to etch out MoS$_2$ NRs using PMMA as a etch mask. Color code used in schematic for each material in the fabrication process is shown in the legend. Figs. f-g show etched NRs in multilayer MoS$_2$ flake with different NR width. Fig. f shows 85 and 185\,nm NRs whereas fig.g shows 50\,nm and 60\,nm NRs.}
\label{fig1} 
\end{figure}

Before fabricating the device, optical characterisation was performed to identify monolayer MoS$_2$ by photoluminiscences and Raman spectroscopy. 
Figs. 2a and 2b show the room temperature photoluminescence measurements (PL) and Raman spectroscopy performed on a flake shown in the  inset of fig.2a.
Both the optical characterisations were performed using a 532 nm laser. 
A PL peak centred at 670\,nm in fig. 2a indicates the presence of direct bandgap, a hallmark of monolayer MoS$_2$.
This peak at 670\,nm corresponds to 1.85eV, a value  consistent with the reported bandgap of monolayer MoS$_2$ \citep{Yee-Fun2018, Splendiani2010}.
For multilayer MoS$_2$ flakes, PL traces (not shown) exhibit multiple emission peaks with weak PL intensities that further reassures that our sample is a monolayer MoS$_2$ flake.
Raman spectroscopy performed using the same laser wavelength (532\,nm) shows the characteristic out-of-plane Raman mode (A$_{1g}$) and the in-plane mode (E$_{2g}$) separated by $\Delta\gamma$=19\,cm$^{-1}$, which agrees well with other reports of  monolayer MoS$_2$ \citep{Yee-Fun2018, Lee2012, Wu2013}.
Devices D1 and D2 were fabricated on a monolayer flake shown in inset of fig. 2a while D3 on another monolayer flake (identified by optical contrast).

All the electrical measurements shown in this work were performed on a Janis probe station at room temperature and under vaccum ($<$10$^{-4}$mTorr) .
Two-terminal source-drain current (I$_d$) was measured by applying a source-drain bias voltage (V) across the device and the  channel carrier density modulated by the backgate (V$_{bg}$).
Figs. 2c and 2d shows the logarithmic source-drain current (I$_d$) traces as a function of V$_{bg}$  for three monolayer MoS$_2$ NR-FET devices (D1-D3) at V=0.1\,V and V=1\,V respectively.
All three devices have a channel of width (W)=50\,nm and channel length (L)=500\,nm.
Typical threshold voltage (V$_{th}$)measured in MoS$_2$ FETs with global back gate are negative values \citep{Banerjee2015}.
We also found negative FET threshold voltages (-30 to -10 V) in all our MoS$_2$ devices after step b) in Fig.1, before the NR etching step.
This is in agreement with the threshold voltages in our MoS$_2$ FETs [-30 to -10\,V] before etching it into NRs. 
After etching the flakes into NRs of width (W)=50\,nm, we see a shift towards positive threshold voltage indicating a transition from depletion mode to accumulation mode in NR FET consistent with previous studies \citep{Liu2012} .
For all the three NR-FET devices (D1-D3), we find a positive threshold voltage in the range 20-35\,V at V=0.1\,V which is extracted using the field effect mobility fit in the linear part of the I$_d$ region (see below for further discussion).
From fig. 2c and 2d, we note that device D1 has slightly higher I$_d$ and steeper sub-threshold slope (SS) compared to other two devices (D2 and D3).
For the rest of the text in this manuscript we focus on the FET characteristics of D1 and for other devices the results are summarize in the table I.

A device exhibiting steep SS allows faster switching between the on- (high current) and the off- (low current) states of FETs.
At V=1\,V, D1 has a SS=3.5\,V/dec.
This value of SS is almost three times steeper then the previous NR-FET value reported of comparable dimensions \citep{Liu2012}.
We achieve steeper SS since our devices are fabricated from single layer MoS$_2$ which provide better electrostatic control over the channel as compared to multilayer channel.
From the SS value, one can extract the interface charge traps using $SS = \frac{2.3k_BT}{q} (1+\frac{C_{it}}{C_{ox}})$ , where k$_B$: Boltzmann constant, T=temperature, q=electron charge,C$_{it}$=interface trap capacitance,C$_{ox}$= gate oxide capacitance \citep{Banerjee2015, Liu2012}.
Using C$_{ox}$(= $\frac{\epsilon_0\epsilon_r}{d}$)=1.15x10$^{-4}$ F/m$^2$ gives interface trap density of D$_{it}$=2.29x10$^{12}$/cm$^2$-eV .
The extracted trap density in our single layer MoS$_2$ FET is better than typical value reported for NR-FET in multilayer devices\citep{Liu2012}.
In order to estimate the SS in a top gated device, we consider a 5\,nm thick hafnium oxide (dielectric constant=25) and assume a same interface trap density to get SS$\approx$69\,mV/dec.
This estimated SS is consistent with the measured SS in single layer MoS${_2}$ FET using hafnium oxide as gate dielectric \citep{Radisavljevic2011}.

Fig. 3a shows the output characteristics of the device D1.
Each I$_d$ vs V trace was taken at a fixed V$_{bg}$ ranging from -20 to 60\,V.
The output characteristic curves are linear at low V exhibiting ohmic behavior, indicating a narrow Schottky barrier.
The curves tend to saturate at higher V due to carrier velocity saturation. 
The saturation I$_d$ gives the on-state current of the FET.
The saturation I$_d$ at V=2\,V and V$_{bg}$=60V is 38$\mu A/\mu m$ giving a on/off ratio of $\approx$10$^5$.
The high current density of 38$\mu$A/$\mu$m in our single layer NR-FET is comparable to the current density in 6\,nm thick NR-FET \citep{Zhang2018} . 
This suggests that our etching process did not degrade the device perfomance significantly. 
The on/off ratio should be considered as the lower limit since we are limited by the off-state current ($\approx$10\,pA) which is 2 orders of magnitude higher than previous report on multilayer NR-FET of similar dimensions \citep{Liu2012,Fathipour2015}.
From the linear region of the output characteristic, field effect mobility can be extracted using the equation $\mu$=$\frac{L}{WC_{ox}V}\frac{dI_d}{dV_{bg}}$ without subtracting the contact resistance \citep{Banerjee2015, Liu2012, Kotekar-Patil2017a}.
Fig. 3b shows the I$_d$ vs V$_{bg}$ at V=0.6V .
Black line overlapping the linear part of I$_d$ in fig.3b is the mobility fit using W=50\,nm and L=500\,nm which gives a mobility of 50 cm$^2$/V.s. 
Fig.3c shows the field effect mobility as a function of V.
We observe that the mobility increases with increase in V and reaches a value of $\mu=50$cm$^2$/V.s at V=0.6\,V.
This increase in mobility with V indicates thinning of Schottky barrier width at higher V resulting in efficient injection of charge carriers.
Comparing the device properties of D1 with the other two devices, D2 and D3 in table 1, we observe variability in transport characteristics.
Even though all the devices are fabricated at the same time with identical process, device characteristics may have variations from device to device depending on the microscopic details of the single layer MoS$_2$, its interface with the substrate, and also the uncertainties introduced by the fabrication process..
\begin{table*}[htbp]
  \centering
  \caption{Table summarizing the NR-FET parameters}
    \begin{tabular}{|c|c|c|c|c|c|c|c|c|}
    \hline 
    Device\# & L(nm) & W(nm) & V$_{th}$(V) & $\mu$(cm$^2$/V.s) & D$_{it}$(10$^{12}$/cm$^2$-eV) & SS(V/dec) & on/off ratio & J($\mu$A/$\mu$m) \\
    \hline 
    D1    & 500   & 50    & 20    & 50    & 2.29  & 3.5   & 1.9x10$^5$ & 38 \\
    D2    & 500   & 50    & 31    & 15.1    & 13.8  & 11.5   & 1.2x10$^5$ & 24 \\
    D3    & 500   & 50    & 33.4    & 31    & 7.8  & 6.5   & 1x10$^5$ & 20 \\
    \hline
    \end{tabular}%
  \label{tab:addlabel}%
\end{table*}%

\begin{figure}
\centering\includegraphics[width=1\columnwidth]{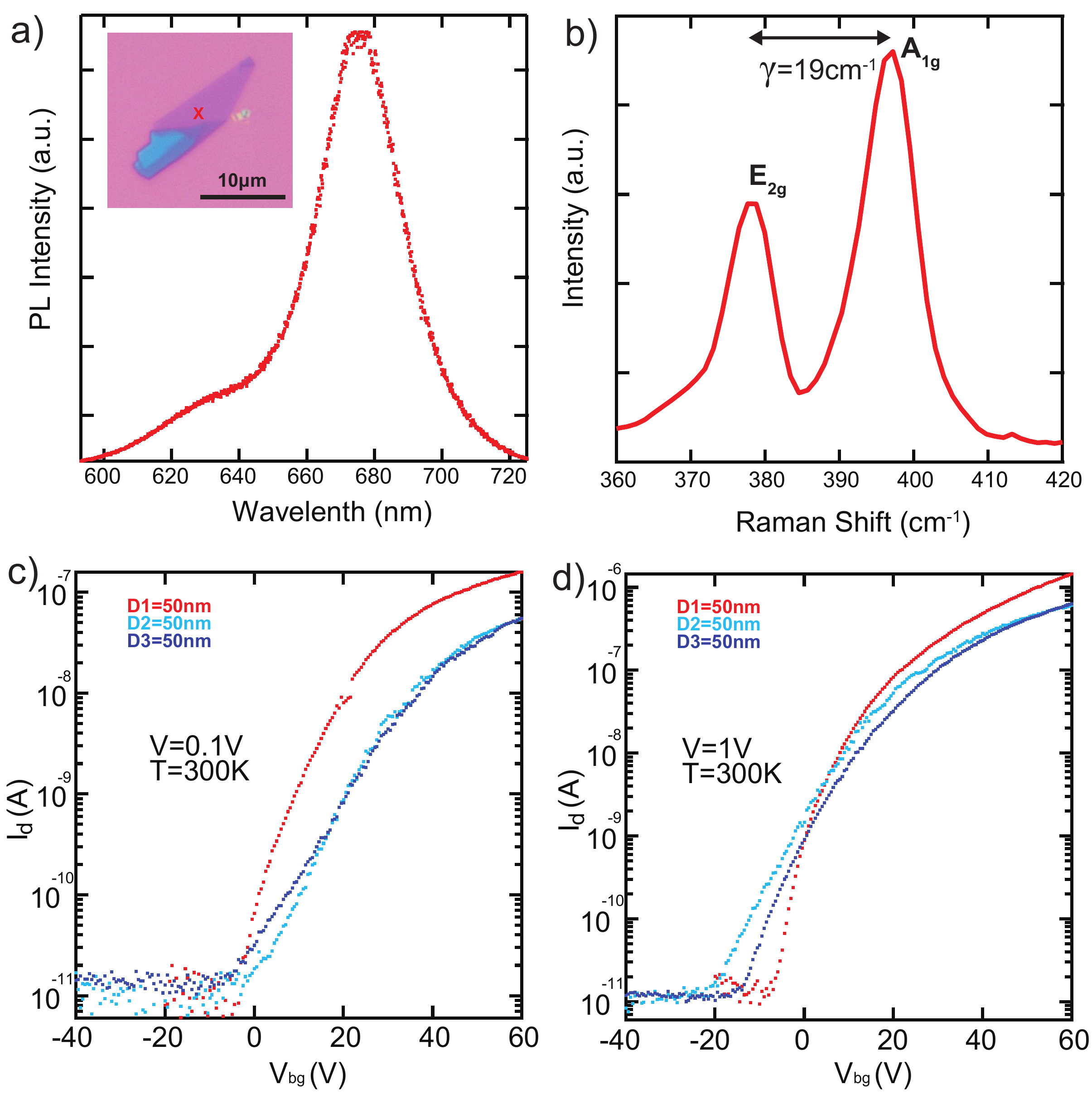}
\caption{a) Room temperature PL measurement showing a strong PL peak at 670\,nm associated with direct bandgap in monolayer MoS$_2$. b) Room temperature Raman spectroscopy showing two peaks at 378.12\,cm$^{-1}$ and 397.18\,cm$^{-1}$ separated by $\Delta \gamma$ =19\,cm$^{-1}$ associated with characteristic E$_{2g} $ and A$_{1g} $ peaks of single layer MoS$_2$ respectively. c) and d) showing logarithmic I$_d$ vs V$_{bg}$ at V=0.1\,V and V=1\,V respectively.} 
\label{fig2} 
\end{figure}

\begin{figure}
\centering\includegraphics[width=1\columnwidth]{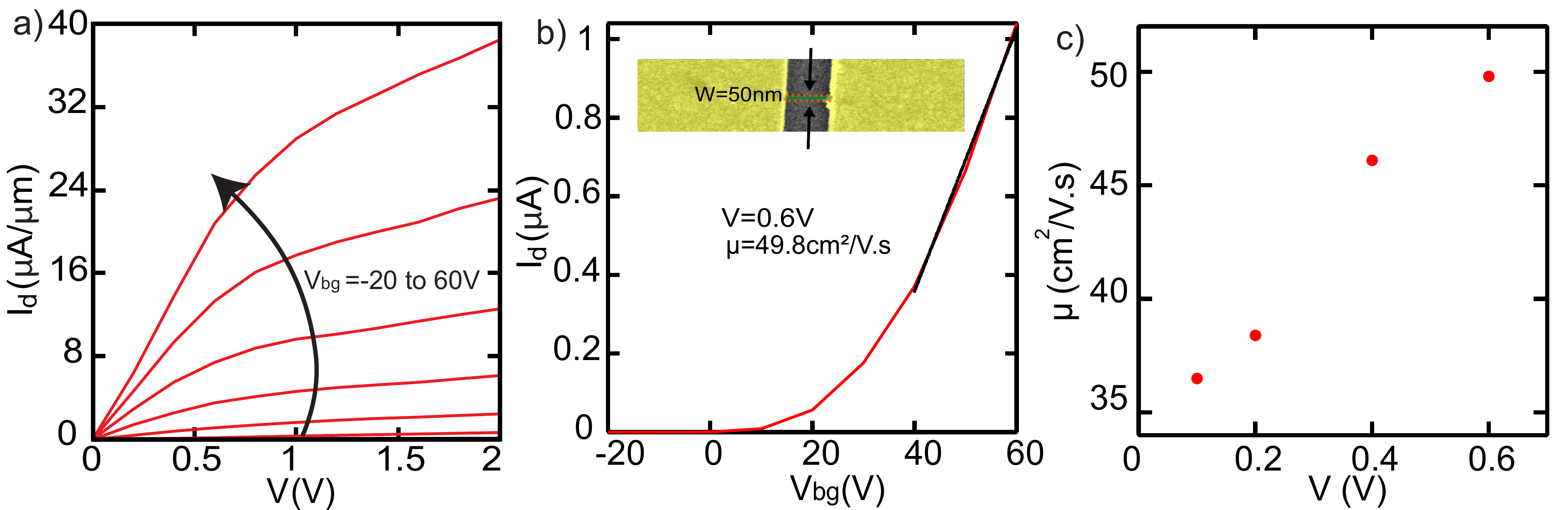} 
\caption{a)  Transfer output characteristic of 50\,nm NR-FET device (D1). Each trace is measured at different back gate voltage varying from V$_{bg}$=-20 to 60\,V. I$_d$ traces increase linearly at low V and approach saturation at high V. b) Linear I$_d$ vs V$_{bg}$ at V=0.6\,V. Black line overlapping the I$_d$ trace is a linear fit to extract mobility. Inset shows false color SEM image of the NR device on which measurement is performed. c) Mobility plotted as a function of V.} 
\label{fig3} 
\end{figure}

To conclude, we have demonstrated a NR-FET etched in a monolayer MoS$_2$  of width 50\,nm with steep sub-threshold slope.
Moreover, we observe increasing mobility with source-drain voltage reaching up to $\mu=50$ cm$^2$/V.s. 
Further improvement in mobility can be  achieved by using a mobility boosting high-$k$ dielectric like HfO$_2$. 
From the transfer characteristic, we find a significant on/off ratio of at least 10$^5$ could be attained even with a back-gated configuration.. 
This work demonstrates a significant milestone in the development of etching monolayer TMDs down to the nm regime, showing that high mobilities in the range of few tens of cm$^2$/V.s is attainable in the first devices fabricated thus.

\section*{Acknowledgments}
This work was supported by the A*STAR (Singapore) QTE Grant No. A1685b0005.

\section*{References}

\bibliography{NR}

\end{document}